\documentclass[aps,prd,floats,twocolumn,twoside,preprintnumbers,superscriptaddress,floatfix,nofootinbib,showpacs]{revtex4-2}
\usepackage{savesym}
\usepackage{latexsym,bbold,colordvi}
\usepackage{graphicx}
\usepackage{amssymb,amsmath,amsthm}
\usepackage{hyperref}
\usepackage{yfonts}
\usepackage{mathrsfs}
\savesymbol{Gray}
\usepackage{siunitx}
\usepackage{slashed}
\usepackage{flushend}
\usepackage[utf8]{inputenc}
\usepackage{verbatim}

\newcommand{\vect}[1]{\mbox{\boldmath $#1$}}
\newcommand{\UMLAUT}[1]{\"{#1}}
\newcommand{\tr}{\mbox{\rm tr}}
\newcommand{\fm}{\mbox{\rm fm}}

\newcommand{\RealNumbers}{\mathbb{R}}
\newcommand{\ComplexNumbers}{\mathbb{C}}
\newcommand{\AEu}{A}
\newcommand{\BEu}{B}
\newcommand{\MEu}{M}
\newcommand{\ZEu}{Z}

\newcommand{\Lqcd}{\Lambda_{\mbox{\rm\scriptsize QCD}}}
\newcommand{\lvac}{\langle\mbox{\rm vac}|}

\newcommand{\bare}{{\it bare}}
\newcommand{\GIA}{GIA}
\newcommand{\BLnonasymptotic}{{\it BL-non-asymptotic}}
\newcommand{\BL}{BL}

\renewcommand{\BibitemShut}[1]{}

\allowdisplaybreaks

\begin{document}

\title{Pion observables calculated in Minkowski and Euclidean spaces with Ans\"{a}tze for quark propagators}

\author{D. Kekez}
\affiliation{Rugjer Bo\v{s}kovi\'{c} Institute, Bijeni\v{c}ka cesta 54,
10000 Zagreb, Croatia}

\author{D. Klabu\v{c}ar}
\affiliation{Physics Department, Faculty of Science,
University of Zagreb, Bijeni\v{c}ka cesta 32, 10000 Zagreb, Croatia}

\date{\today}

\preprint{ZTF-EP-20-03}

\begin{abstract}
\noindent
We study two quark--propagator meromorphic {\it Ans\"atze} that admit clear
connection between calculations in Euclidean space and Minkowski spacetime.
The connection is established through a modified Wick rotation in momentum space,
where the integration contour along the imaginary axis is adequately deformed.
The {\it Ans\"atze} were previously proposed in the literature and
fitted to Euclidean lattice QCD data.
The generalized impulse approximation is used to calculate the pion transition
form factor and electromagnetic form factor, correcting an earlier result.
The pion decay constant and distribution amplitude are also calculated.
The latter is used to deduce the asymptotic behavior of the form factors.
Such an asymptotic behavior is compared
with those  obtained directly from the generalized impulse approximation
and the causes of differences are pointed out.
\end{abstract}

\maketitle

\section{Introduction}
\label{sec:Introduction}

  Obtaining the properties of hadrons as quark and gluon bound states,
from the underlying theory of strong interactions, QCD, has proven to be
extremely challenging. Reproducing even relatively simple observables,
such as decay constants, is 
difficult whenever the nonperturbative regime of QCD must be dealt with.
However, powerful tools for this task have been developed over the last decades.
These tools include lattice QCD \cite{LQCD2020,USQCD:2022mmc} 
calculations in Euclidean space
and continuum functional methods. The latter is exemplified by
functional renormalization group
(see, {\it e.g.}, Refs.~\cite{Pawlowski:2005xe,Schaefer:2006sr} and references
therein)
and Schwinger--Dyson equations (SDE); see, {\it e.g.}, 
Refs.~\cite{Alkofer:2000wg,Roberts:2000aa,Fischer:2006ub,Roberts:2020hiw} for reviews and
Refs.~\cite{Kekez:1996np,Klabucar:1997zi,Kekez:1998xr,Kekez:1998rw,Kekez:2001ph,Kekez:2003ri,Kekez:2005ie}
for examples of calculations of some observables addressed
also in the present paper.
A general discussion about meson physics in new experimental programs
is provided by Refs.~\cite{Dudek:2012vr,Aguilar:2019teb}.

  Due to technical complications inherent to these two continuum functional
approaches, most corresponding calculations are not done in physical
Min\-kow\-ski spacetime but, again, in four-dimensional Euclidean space.
Hereby one exploits a technical trick, the so-called Wick rotation, to map
quantum field theory in Minkowski spacetime to Euclidean space.
The situation with the Wick rotation relating Minkowski with Euclidean space
must be under control, but this is highly nontrivial
in the nonperturbative case.
In particular, it should be clarified whether nonperturbative QCD Green's
functions employed in a calculation permit Wick rotation.
In this work, we do it for two strongly dressed quark-propagator
{\it Ans\"atze} \cite{Mello:2017mor,Alkofer:2003jj} modeling nonperturbative QCD.

   On the formal level, the Osterwalder--Schrader reconstruction theorem
states that the Schwinger functions of some Euclidean
field theory can be analytically extended to Wightman functions of
the corresponding Minkowski space quantum field theory,
providing that these Schwinger functions satisfy some set of constraints,
the Osterwalder--Schrader axioms \cite{Osterwalder:1973dx}.

 The widely used rainbow--ladder truncation to the coupled SDE
for the dressed quark propagator (``gap equation'') and
Bethe--Salpeter Equation (BSE) for a quark--antiquark
bound state are usually formulated in the Euclidean
space and equations are solved for spacelike momenta
\cite{Roberts:1994dr}.
Although some physical quantities
can be extracted from the results in Euclidean space alone, many others,
such as, {\it e.g.}, decay properties,
cannot be calculated with just real Euclidean four--momenta.
In general, for solving the BSE and calculation
of processes, knowledge is needed about the analytic behavior
in part of the complex momentum--squared plane
(see, {\it e.g.}, Ref.~\cite{Alkofer:2002bp}).
In this respect, analytic continuation of auxiliary quantities like
Green's functions of the theory, notably the quark propagator, opens up
the possibility to provide an understanding of strong-interaction processes
from results of lattice QCD and functional methods.
   
The use of such analytically continued propagators should be tried 
in calculations of hadron observables from the QCD substructure. The
pion decay constant is an example of a relatively simple such quantity,
whereas the pion form factors are already on a much higher level of
difficulty; namely, due to their momentum dependence, one must take into
consideration both the perturbative and nonperturbative regime of QCD.
The charged pion electromagnetic form factor (EMFF) is calculated
to next--to--nexto--to--leading order in chiral perturbation theory \cite{Bijnens:1998fm},
using the QCD sum rules \cite{Ioffe:1982ia},
vector--meson dominance \cite{Dominguez:1982xb},
Sudakov suppression \cite{Li:1992nu},
light--cone sum rules \cite{Braun:1994ij,Braun:1999uj,Bijnens:2002mg},
AdS/QCD correspondence \cite{Grigoryan:2007wn,Brodsky:2007hb},
lattice QCD in quenched approximation \cite{Martinelli:1987bh,Draper:1988bp},
or with the dynamical quarks \cite{Brommel:2006ww,Frezzotti:2008dr,Bonnet:2004fr,Koponen:2015tkr}
The transition form factor (TFF) is calculated using the
QCD sum rules \cite{Radyushkin:1995pm,Radyushkin:1996tb},
light--cone sum rules
\cite{Mikhailov:2009kf,Stefanis:2012yw,Mikhailov:2016klg,Stefanis:2020rnd},
light--front constituent quark models
\cite{El-Bennich:2012mkr,deMelo:2013zza,Choi:2016jot,Choi:2017zxn,Choi:2019wqx},
vector--meson dominance \cite{Kessler:1993gz},
anomaly sum rule \cite{Klopot:2010ke,Klopot:2012hd},
Sudakov suppression \cite{Jakob:1994hd,Ong:1995gs,Stefanis:2000vd},
lattice QCD \cite{Gerardin:2016cqj,Gerardin:2019vio}, and
large--$N_c$ chiral perturbation theory \cite{Bickert:2020kbn}.

  However, only a limited number of papers deal with the quark--propagator
modeling, or solving its SDE, in Min\-kow\-ski space.
\v{S}auli, Adam, and Bicudo \cite{Sauli:2006ba} have explored
the fermion--propagator SDE in Minkowski space.
The interaction used is a meromorphic
function of momentum transfer squared; it has two simple poles on the
real axis, in the timelike region.
Various spectral representations of the fermion propagator are employed.
Ruiz Arriola and Broniowski \cite{RuizArriola:2003bs} have proposed
a spectral quark model based on a generalization of the Lehmann representation
of the quark propagator and applied it
to calculate some low--energy quantities.
While their $\sigma_V$ and $\sigma_S$ functions
[defined by Eq.~(\ref{QuarkPropagatorParametrization})]
exhibit only cuts on the timelike part of the real axis, 
the quark dressing function $\AEu(z)$
[see Eq.~(\ref{QuarkPropagatorParametrization})] has pairs of the
complex--conjugate poles in the complex momentum plane.
{}Siringo \cite{Siringo:2015wtx,Siringo:2016jrc}
has studied the analytic properties of gluon, ghost, and quark propagators
in QCD, using a one--loop massive expansion in the Landau gauge.
He studies spectral functions in Minkowski space, by analytic continuation
from deep infrared, and finds complex conjugated poles for the gluon propagator
but no complex poles for the quark propagator.
{}A group of interconnected papers
\cite{Roberts:1994hh,Frank:1994gc,Roberts:2010rn,Chang:2013nia,Mezrag:2014jka,
Raya:2015gva,Horn:2016rip} typically start from a consistently truncated system
of SDE and BSE, or some algebraic
{\it Ans\"atze} for the quark propagator and Bethe--Salpeter (BS) amplitude 
inspired by such a consistent system.
They have calculated the EMFF, TFF, and pion distribution amplitude (PDA),
sometimes relying on Nakanishi--like representation
\cite{Nakanishi:1963zz,Nakanishi:1969ph,nakanishi1971graph}
to solve the practical problem of continuing from Euclidean space to
Minkowski space \cite{Chang:2013nia}.
The Nakanishi representation is also used in Refs. \cite{Mezrag:2020iuo,Duarte:2022yur}.
The covariant spectator theory is related to the SDE and BSE in Minkowski space
\cite{Gross:1969rv,Biernat:2013fka,Biernat:2014xaa,Biernat:2018khd}; one starts
with the usual BSE with one particle restricted to the mass shell, resulting
in a three--dimensional equation. In addition to the one--gluon exchange, the
interaction kernel may include a covariant generalization of linear confining potential.
The pion EMFF is calculated in Refs.~\cite{Biernat:2013aka,Biernat:2015xya,Biernat:2018tlj}.

   In this work we study two quark--propagator {\it Ans\"{a}tze}.
The first one is by
Mello, de Melo, and Frederico (MMF) \cite{Mello:2017mor},
and the second on by Alkofer, Detmold, Fischer, and Maris (ADFM)
\cite{Alkofer:2003jj}.
The propagators are defined in momentum space;
the pertinent dressing functions are meromorphic functions of momentum squared,
exhibiting only simple poles on the timelike part of the real axis.
On the good side,
such a simple analytic structure makes the Wick rotation allowed
and technically feasible, at least for the processes and approximation
schemes under consideration.
The {\it Ans\"{a}tze} are fitted to the lattice data,
which are available for the spacelike  momenta.
On the bad side, the meromorphic {\it Ans\"{a}tze} are not able to reproduce
the perturbative QCD (pQCD) asymptotic behavior, and
we showed that this deficiency impairs
calculation of some processes, notably the high--$Q^2$ behavior of the
form factors. In the present work these {\it Ans\"{a}tze} are used to obtain
the pion decay constant, neutral pion TFF, charged pion EMFF,
and PDA. In particular, we correct the result for the pion EMFF
given in Ref.~\cite{Mello:2017mor}.

  The remainder of the paper is organized as follows.
Sections \ref{sec:MelloQP} and \ref{sec:ThreeRQP} introduce
the quark--propagator models of Refs.~\cite{Mello:2017mor} and
\cite{Alkofer:2003jj}, respectively.
In Sec.~\ref{sec:PionDecayConstant}, the pion decay constant is calculated;
approximation and numerical methods, which will be used throughout
the paper, are presented. In Sec.~\ref{sec:EFF}, the pion EMFF
is calculated, while Sec.~\ref{sec:TFF} deals with
the TFF.
The calculation of the PDA is addressed in
Sec.~\ref{sec:PDA} and the obtained distribution is used to calculate
the asymptotic form of the TFF.
Various approximations are investigated and compared with those of
Secs. \ref{sec:EFF} and \ref{sec:TFF}. 
Section \ref{sec:SummaryAndConclusions} provides summary and conclusions.


\section{MMF quark propagator}
\label{sec:MelloQP}

  The dressed quark propagator in a general covariant gauge can be written as
\begin{align}
&S(q) = \ZEu(-q^2) [\slashed{q} - \MEu(-q^2)]^{-1}
\nonumber \\
&=[\AEu(-q^2) \slashed{q} - \BEu(-q^2)]^{-1}
\nonumber \\
& = - \sigma_V(-q^2) \slashed{q} - \sigma_S(-q^2)~,
\label{QuarkPropagatorParametrization}
\end{align}
%
where $\MEu=\BEu/\AEu$ is the renormalization--point independent
quark mass function and $\ZEu=1/\AEu$ is the wave function renormalization
(see, {\it e.g.}, Ref.~\cite{Roberts:1994dr}).
The Minkowski metric is used, with the signature
$(\begin{array}{cccc}+&-&-&-\end{array})$.
The MMF quark propagator~\cite{Mello:2017mor}
is fixed by the following quark mass function and wave function
renormalization parametrization:
\begin{subequations}
\label{QuarkPropagators:MelloMeloFredericoQP:MandA}
\begin{align}
\MEu(x) & = (m_0-i\varepsilon)
  + m^3 \left[ x + \lambda^2 - i \varepsilon \right]^{-1} ~,
\label{QuarkPropagators:MelloMeloFredericoQP:MEu}
\\
\ZEu(x) &= 1~,
\label{QuarkPropagators:MelloMeloFredericoQP:AEu}
\end{align}
\end{subequations}
where $m_0=0.014~\unit{\GeV}$, $m=0.574~\unit{\GeV}$, and $\lambda=0.846~\unit{\GeV}$.
The infinitesimally small parameter $\varepsilon$ prescribes 
how to treat contour integration around poles.
The function $\MEu$ is shown as the blue solid line in Fig.~\ref{fig:Mello:lattice}.
(This {\it Ansatz} form has been already used to fit lattice QCD
data \cite{Dudal:2013vha}. There, the parameter values $m_0$, $m$,
and $\lambda$ are rather close to those used in Ref.~\cite{Mello:2017mor}
and in the present paper; nevertheless, 
the propagator of Ref.~\cite{Dudal:2013vha} exhibits one real
and a pair of complex--conjugated poles.)
Asymptotic expansions of $\MEu$ about $\infty$ and $0$ are
\begin{align}
\MEu(x) &= m_0 + \frac{m^3}{x} - \frac{\lambda^2 m^3}{x^2} 
  + {\cal O}((\frac{1}{x})^3)~,
\label{MEinf:Mello}
\\
\MEu(x) &= \left( m_0 + \frac{m^3}{\lambda^2} \right)
        - \frac{m^3x}{\lambda^4} + \frac{m^3 x^2}{\lambda^6}
        + {\cal O}(x^3)~,
\end{align}
respectively.
The functions $\AEu$, $\BEu$, $\sigma_V$, and $\sigma_S$
depend algebraically on $\ZEu$ and $\MEu$ and are defined for convenience.
The quark dressing functions $\sigma_V$ and $\sigma_S$,
introduced by Eq.~(\ref{QuarkPropagatorParametrization}),
can be decomposed as
\begin{subequations}
\label{MMFQP}
\begin{align}
\sigma_V(x) &= \sum_{j=1}^3 \frac{b_{Vj}}{x+\mathfrak{p}_j}~,
\\
\sigma_S(x) &= \sum_{j=1}^3 \frac{b_{Sj}}{x+\mathfrak{p}_j}~,
\end{align}
\end{subequations}
%
where the coefficients $\mathfrak{p}_j$, $b_{Vj}$, and $b_{Sj}$ $(j=1,2,3)$
are certain complicated algebraic functions of the parameters
$m_0$, $m$, and $\lambda$.
Obviously, $\sigma_{V,S}(x) \to 0$ for all $x \to \infty$.

\section{ADFM quark propagator}
\label{sec:ThreeRQP}

\begin{figure}
\centerline{\includegraphics[width=8.5cm,angle=0]{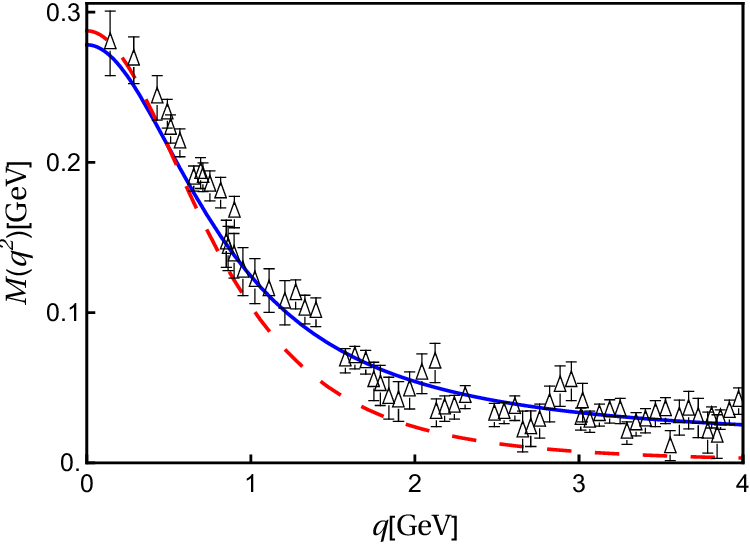}}
\caption{Blue solid line and red dashed line correspond to
the MMF{} and ADFM quark--propagator{} {\it Ans\"{a}tze}, respectively.
Lattice data \cite{Parappilly:2005ei} are represented by the open triangles.}
\label{fig:Mello:lattice}
\end{figure}

 The dressing functions $\sigma$ of the ADFM meromorphic
 {\it Ansatz} \cite{Alkofer:2003jj} that have three real poles
 (by choosing {\it their} $b_j=0$, see Ref. \cite{Alkofer:2003jj}) are
\begin{subequations}
\label{3RQP}
\begin{eqnarray}
\sigma_V(x) &=& \frac{1}{Z_2}
   \sum_{j=1}^3
   \frac{2 r_j}
        { x + a_j^2 }~,
\\
\sigma_S(x) &=& \frac{1}{Z_2}
   \sum_{j=1}^3
   \frac{2 r_j a_j }
        { x + a_j^2  }~,
\end{eqnarray}
\end{subequations}
where
$a_1=0.341~\unit{\GeV}$, $a_2=-1.31~\unit{\GeV}$, $a_3=-1.35919~\unit{\GeV}$,
$r_1=0.365$, $r_2=1.2$, $r_3=-1.065$,
$Z_2=0.982731$ \cite{Alkofer:2003jj}.
The coefficients $r_j$ and $a_j$ satisfy
\begin{align}
\sum_{j=1}^{3} r_j = \frac{1}{2}~, & \qquad\qquad
\sum_{j=1}^{3} a_j r_j = 0~.
\end{align}
The first of the above constraints follows from the consideration of the
large--momentum limit of $\sigma_V(x)$; the second one arises from
the requirement that $\MEu(x)$ must vanish for large spacelike real momenta.
\footnote{Away from the chiral limit, the second sum would be equal
to the renormalized quark mass.}
The {\it Ansatz} (\ref{3RQP}) guarantees that the quark dressing functions
$\sigma_{S,V}(z)\to 0$ for all $|z|\to\infty$
in the complex $z$ plane \cite{Oehme:1996ju}.
For the given set of parameters the functions
$x\mapsto \AEu(-x)$
and $x\mapsto \BEu(-x)$ have two real poles for $x<0$
(see ${\mathfrak b}_{1,2}$ below Eq. (\ref{Eta4Pi:Bdecomposition})).
The corresponding quark mass function $\MEu$ is shown 
as the red dashed line in Fig.~\ref{fig:Mello:lattice}.

%

   Euclidean formalism adopted in Ref.~\cite{Alkofer:2003jj} avoids
probation of the quark dressing functions (\ref{3RQP}) near their poles,
$x=-a_j^2$, $j=1,2,3$. As we want to analytically continue $\sigma$'s to
the complex plane and use these functions for the
calculation in Minkowski space,
a prescription for the pole treatment ought to be defined.
An obvious choice is Feynman's $i\varepsilon$ prescription,
already used in the MMF--{\it Ansatz} case
(\ref{QuarkPropagators:MelloMeloFredericoQP:MEu});
we push the poles  infinitesimally from the real axis:
$x=-a_j^2+i\varepsilon$, $j=1,2,3$. We use this prescription throughout
this paper.

   Functions $\AEu(x)$ and $\BEu(x)$ that follow from Eqs.~(\ref{3RQP})
are also rational functions, exhibiting real poles for $x<0$.
For example, Eqs.~(\ref{QuarkPropagatorParametrization}) imply that
function $\BEu$, which will be used
in further calculation, is
\begin{align}
\BEu(x) &= \frac{\sigma_S(x)}{\sigma_S^2(x) + x\, \sigma_V^2(x)}
\nonumber \\
    &= - \frac{\mathfrak c}{{\mathfrak b}_1-{\mathfrak b}_2}
		     \left[
        \frac{({\mathfrak b}_1-{\mathfrak a})}{(x+{\mathfrak b}_1)}
      + \frac{({\mathfrak a}-{\mathfrak b}_2)}{(x+{\mathfrak b}_2)} \right]~,
\label{Eta4Pi:Bdecomposition}
\end{align}
%
where the coefficients ${\mathfrak a}$, ${\mathfrak b}_1$,
${\mathfrak b}_2$, and ${\mathfrak c}$ are some complicated algebraic
functions of the original parameters $Z_2$, $a_j$, and $r_j$, appearing
in Eqs.~(\ref{3RQP}).
Their calculated values are
${\mathfrak a}=38.1104~\unit{\GeV}^2$,
${\mathfrak b}_1=0.488784~\unit{\GeV}^2$,
${\mathfrak b}_2=2.65383~\unit{\GeV}^2$, and
${\mathfrak c}=-0.0178316~\unit{\GeV}^3$.
A small $i\varepsilon$ shift of $\sigma_V$ and $\sigma_S$
poles, $x=-a_j^2+i\varepsilon$, $j=1,2,3$,
causes the similar shift of the $\BEu$ poles,
$x=-{\mathfrak b}_k+i\varepsilon^\prime$, $k=1,2$,
in agreement with the Feynman prescription.

   For $z\in\ComplexNumbers$ and large  $|z|$ we find that $\MEu(z)\propto 1/z$,
but this asymptotic behavior is reached only at very high momenta squared,
$|z|\simeq 1000~\unit{\GeV}^2$. The MMF quark--propagator {\it Ansatz} shows the same asymptotics
for $m_0=0$, while $\MEu(z)\sim m_0$ for $m_0\ne 0$;
see Eq.~(\ref{MEinf:Mello}).
A well--known QCD result \cite{Lane:1974he,Politzer:1976tv} for the
asymptotics of the quark mass function is
\begin{equation}
\MEu(z)\propto
\left\{
\begin{array}{cc}
{}[\log(z/\Lqcd^2)]^{d-1}/z & \mbox{in the chiral limit}
\\
{}[\log(z/\Lqcd^2)]^{-d}   & \mbox{otherwise}
\end{array}
\right.~,
\label{MEu:asymptotics}
\end{equation}
where $d=12/(11N_c-2N_f)$ is the anomalous mass dimension,
$N_c$ and $N_f$ are the number of colors and flavors, respectively and
$\Lqcd\sim 0.5~\unit{\GeV}$ is the QCD scale.
The simple meromorphic {\it Ans\"atze},
Eqs.~(\ref{3RQP}) and (\ref{QuarkPropagators:MelloMeloFredericoQP:MandA}),
emulate the chiral--limit and away--from--the--chiral--limit behavior,
respectively, of the quark mass function (\ref{MEu:asymptotics}),
up to the logarithmic corrections present in Eq.~(\ref{MEu:asymptotics}).
The {\it Ans\"atze} are fitted to the respective lattice data:
{MMF quark propagator} to lattice data of Ref.~\cite{Parappilly:2005ei} and
{ADFM quark propagator} to lattice data
in the overlap \cite{Bonnet:2002ih,Zhang:2002bg,Zhang:2003faa}
and asqtad (tadpole improved staggered) \cite{Bowman:2002bm} formulations.

\section{Pion decay constant}
\label{sec:PionDecayConstant}

\begin{figure}[t]
\begin{center}
\includegraphics[width=9cm,angle=0]{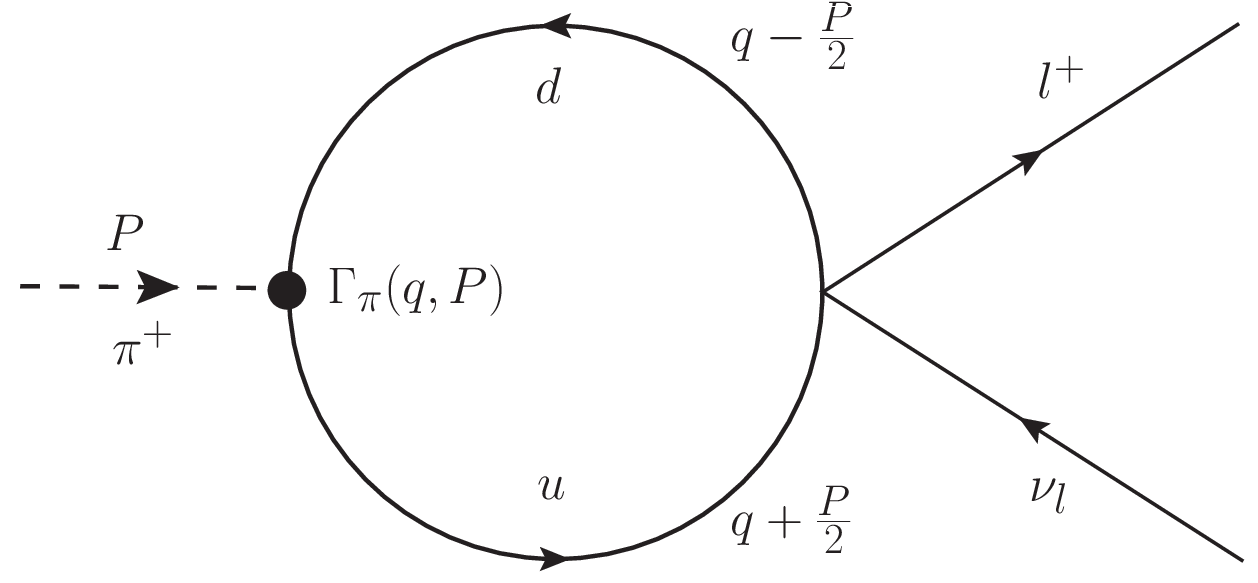}
\end{center}
\caption{Diagram for $\pi^+\to l^+\nu_l$ decay.}
\label{fig:PionDecayConstant}
\end{figure}

The pion decay constant $f_\pi$ is defined by the matrix element
\begin{equation}
\langle 0| \bar{d}(x) \gamma^\mu \gamma_5 u(x) |\pi^{+}(P)\rangle =
i \sqrt{2} f_{\pi} P^\mu e^{-iP\cdot x}~,
\label{MesonDecayConstant:fpip}
\end{equation}
%
where $u(x)$ and $d(x)$ are the quark fields
(see, {\it e.g.}, Ref.~\cite{ParticleDataGroup:2020ssz}, Sec.~71.1).
This matrix element is the hadronic part of the amplitude
for $\pi^+\to l^+\nu_l$ decay, pictorially represented in
Fig.~\ref{fig:PionDecayConstant}.
More explicitly,
$f_\pi$ can be expressed in terms of the BS
vertex function $\Gamma_\pi(q,P)$,
\begin{align}
f_\pi &= i \frac{N_c}{2 M_\pi^2} 
	\nonumber \\ & \times
        \int \frac{d^4q}{(2\pi)^4}
        \mbox{\rm tr}
        \left( \slashed{P} \gamma_5 S(q+\frac{P}{2})
		\Gamma_\pi(q,P)
                S(q-\frac{P}{2}) \right)~,
\label{fpi_approx}
\end{align}
where $N_c=3$ is the number of colors, and $M_\pi$ is the pion mass.
Dictated by dynamical chiral symmetry breaking
the axial--vector Ward--Takahashi identity, taken in the chiral limit, gives us
the quark--level Goldberger--Treiman relation for the BS vertex,
\begin{equation}
\Gamma_\pi(q,P) \simeq  -\frac{2 \BEu(-q^2)_{\mbox{\rm\scriptsize c.l.}}}{f_\pi}\, \gamma_5~,
\label{GTrelation}
\end{equation}
which expresses $\Gamma_\pi$ in terms of the chiral--limit (c.l.) value of the
quark dressing function $\BEu$; see, {\it e.g.}, Ref.~\cite{Roberts:1994dr}.
This approximation will be used throughout this paper. [Note that it is the
same approximation as in Ref.~\cite{Mello:2017mor}, as can be seen easily
in spite of different notations and conventions, by comparing their Eqs. (18),
(19) and (21) with our Eqs. (\ref{GTrelation}) and (\ref{Gamma-parametrization}).
See also our Appendix.]

The pion decay constant $f_\pi$ corresponding to
the MMF quark--propagator model
(\ref{QuarkPropagators:MelloMeloFredericoQP:MandA})
has been calculated in three different ways:
(a) analytically using {\it Mathematica} packages
FeynCalc 9.0 \cite{Mertig:1990an,Shtabovenko:2016sxi}
and {\footnotesize PACKAGE-X} 2.0 \cite{Patel:2015tea,Patel:2016fam},
(b) numerical integration in the Euclidean space, and
(c) Minkowski space integration utilizing light-cone momenta and
analytic residua calculation.
Let us explain them in more detail.

   (a) Using FeynCalc it is possible to express $f_\pi$
as a sum of terms containing
Passarino-Veltman functions $B_0$ \cite{Passarino:1978jh} of various arguments.
{\footnotesize PACKAGE-X} is subsequently used for the
final numerical evaluation, giving $f_\pi=87.5599~\unit{\MeV}$. The same result
is obtained using LoopTools 2.0 \cite{Hahn:1998yk}
for the final numerical evaluation.

   (b) The naive prescription for Wick rotation
($q^0\to -i q^4$, $\int dq^0\to i \int dq^4$)
is justified here,
for this specific propagator and for the pion decay constant calculation.
Numerical integration in Euclidean space gives again the same
$f_\pi$, to at least six significant digits. The four--dimensional
integration is effectively two--dimensional, two integrations are
trivial due to symmetry. The pion mass is taken to be $M_\pi=135~\unit{\MeV}$.

   (c) Alternatively, following
the procedure used in Ref.~\cite{Mello:2017mor},
integral (\ref{fpi_approx}) is calculated introducing
light--cone variables $q_{\pm} = q^0 \pm q^3$.
The integrand is a rational function in $q_-$ variable, with seven simple
poles on the real $q_-$ axis. Cauchy's residue theorem is used to
calculate the integral over $q_-$, paying attention to the $i\varepsilon$
rule for the displacement of poles,  prescribed by
Eq.~(\ref{QuarkPropagators:MelloMeloFredericoQP:MEu}).
The remaining two--dimensional integration over $q_+\in[-M_\pi/2,M_\pi/2]$ and
$(q^1)^2+(q^2)^2$ is performed numerically. Eventually, the resulting
$f_\pi=87.5599~\unit{\MeV}$ is in agreement with our previous calculations.
The result of Ref.~\cite{Mello:2017mor} is $f_\pi=90~\unit{\MeV}$,
a little above our calculated value.

   Regarding the ADFM {\it Ansatz}, $f_\pi$ is calculated
using methods (a) and (b) mentioned above,
and (d). Method (d) is the
Minkowski space integration where the first integration, over $q^0$,
boils down to residua calculation, as the principal value vanishes.
All three methods give the same result, $f_\pi = 71.5611~\unit{\MeV}$.
Regarding the method (a),
the trace appearing in Eq.~(\ref{fpi_approx}) is evaluated
using FeynCalc and LoopTools {\it Mathematica}
packages, formally treating
$\BEu(x)$ as a sum of two propagators [see Eq.~(\ref{Eta4Pi:Bdecomposition})].

\section{Electromagnetic form factor}
\label{sec:EFF}

\begin{figure}[t]
\centerline{\includegraphics[width=9cm,angle=0]{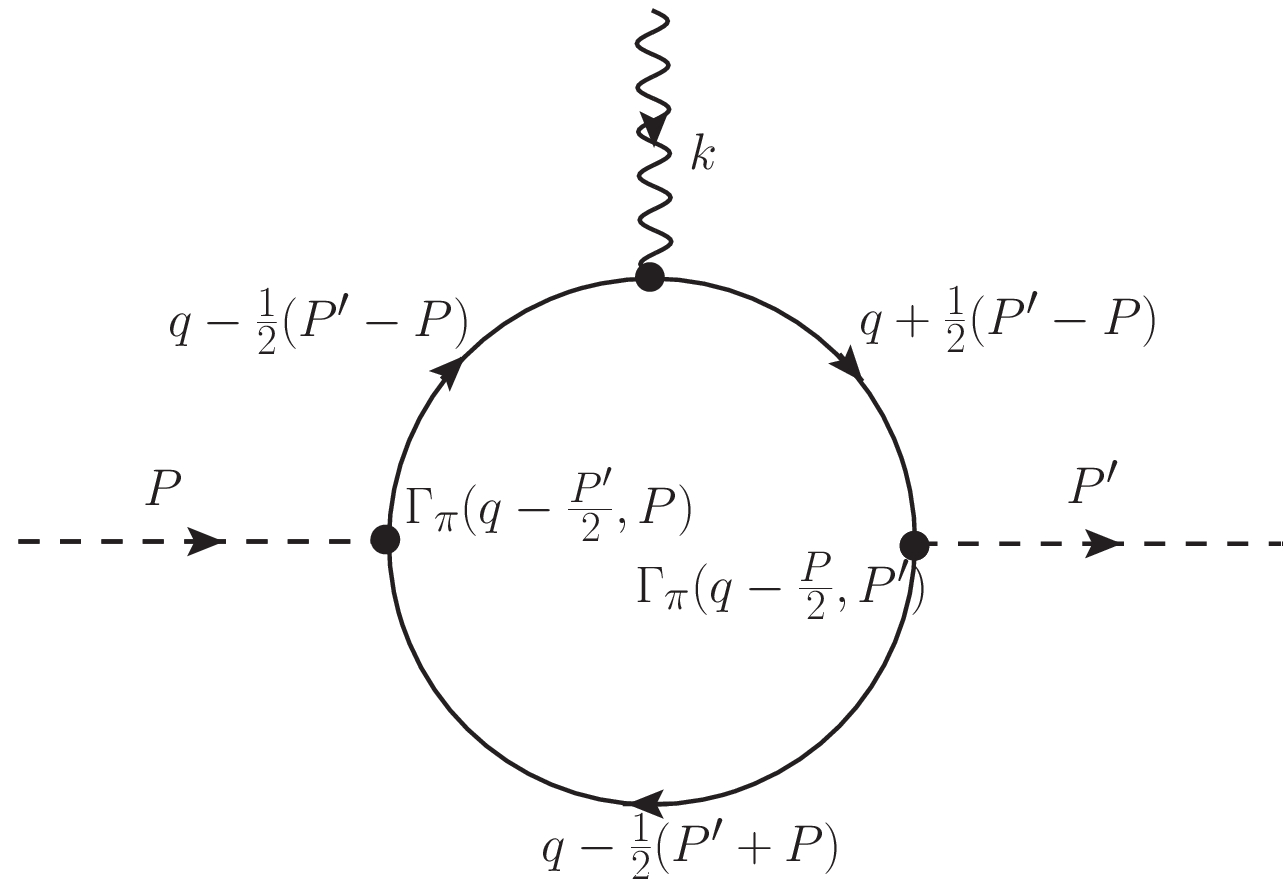}}
\caption{Impulse approximation to
the charged pion electromagnetic form factor $F_{\pi}(Q^2)$.}
\label{QuarkPropagators:pion-emff}
\end{figure}

   The charged pion EMFF $F_\pi(Q^2)$ is given by 
\begin{align}
&\langle \pi^{+}(P^\prime)|J^\mu(0)|\pi^{+}(P)\rangle
=
\mathcal{Q}_{\pi^{+}} (P^\mu + P^{\prime\mu}) F_\pi (Q^2) \nonumber \\
&=
i (\mathcal{Q}_u-\mathcal{Q}_d) \frac{N_c}{2}
\int \frac{d^4q}{(2\pi)^4}
\mbox{\rm tr} 
	\Bigl\{
	\bar{\Gamma}_\pi(q-\frac{P}{2},P^\prime)
	\nonumber \\ &\times
        S(q+\frac{1}{2}(P^\prime-P))
	\Gamma^\mu(q+\frac{1}{2}(P^\prime-P),q-\frac{1}{2}(P^\prime-P))
        \nonumber \\ &\times
        S(q-\frac{1}{2}(P^\prime-P))
	\Gamma_\pi(q-\frac{1}{2}P^\prime,P)
	S(q-\frac{1}{2}(P+P^\prime))
	\Bigr\}~,
\label{Pi-EM-form-factor(4)}
\end{align}
%
in the generalized impulse approximation (GIA)
\cite{Pagels:1979hd,Roberts:1993ks,Alkofer:1993gu}, for spacelike $Q^2$,
and the momentum routing as depicted in Fig.~\ref{QuarkPropagators:pion-emff}.
The electromagnetic current is
$J^\mu(x)$;
the quark charge $\mathcal{Q}_u=2/3$ and $\mathcal{Q}_d=-1/3$.
We use the following kinematics:
$k=(0,0,0,\sqrt{Q^2})$,
$P=(E_\pi,0,0,-{\sqrt{Q^2}}/{2})$,
and
$P^\prime=(E_\pi,0,0,\allowbreak{\sqrt{Q^2}}/{2})$,
where $E_\pi=\sqrt{M_\pi^2+{Q^2}/{4}}$ and $Q^2\ge 0$.
The Ball--Chiu vertex \cite{Ball:1980ay,Ball:1980ax} is used for
the quark--quark--photon coupling throughout this paper,
\begin{align}
\Gamma^\mu(p^\prime,p)
= \frac{1}{2}
[  \AEu(-p^{\prime 2}) + \AEu(-p^2) ] \gamma^\mu
+ \frac{ (p^\prime+p)^\mu } { ( p^{\prime 2} - p^2) }
\nonumber \\ \times
\Bigr\{ [ \AEu(-p^{\prime 2}) - \AEu(-p^2) ]
\frac{\textstyle ( \slashed{p}^\prime + \slashed{p} ) }{2}
-[ \BEu(-p^{\prime 2}) - \BEu(-p^2) ] \Bigr\}~.
\label{Ball-Chiu-vertex}
\end{align}
This vertex can be expressed completely in terms of the quark--propagator
dressing functions
and it becomes particularly simple in the case of the MMF{} {\it Ansatz}:
\begin{equation}
\Gamma^\mu(p^\prime,p)
=
\gamma^\mu
-
\frac{m^3 (p^{\prime\mu}+p^\mu)}
     {(p^{\prime 2}-\lambda^2+i\varepsilon)(p^2-\lambda^2+i\varepsilon)}~.
\end{equation}

Similarly to the case of $f_\pi$ calculation,
three methods are used to calculate $F_\pi(Q^2)$
using the MMF{} {\it Ansatz}:
(a) FeynCalc and {\footnotesize PACKAGE-X} {\it Mathematica} packages,
(b) numerical integration in Euclidean space using adaptive quadrature, and
(c) Minkowski space integration utilizing light--cone momenta
and analytic residua calculation.
Let us discuss these methods in more detail.
  
\begin{figure}[t]
\centerline{\includegraphics[width=9cm,angle=0]{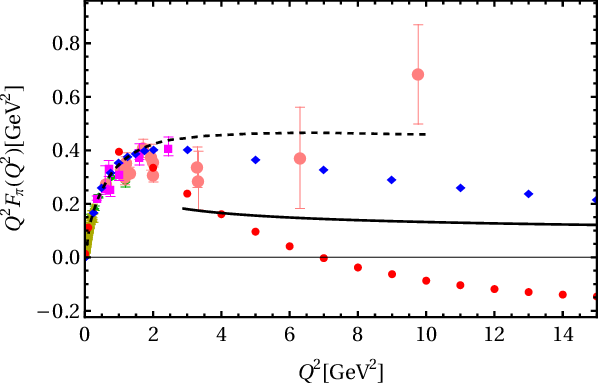}}
\caption[Caption for LOF]{Charged pion electromagnetic form factor.
Experimental points are shown by
dark yellow triangles \cite{Amendolia:1986wj},
green diamonds \cite{Brown:1973wr},
pink circles \cite{Bebek:1974iz,Bebek:1974ww,Bebek:1977pe}, and
magenta squares \cite{Ackermann:1977rp,Brauel:1979zk,Blok:2008jy,Huber:2008id}.
Red solid circles and blue diamonds
are calculated using the ADFM quark--propagator{} {\it Ansatz}
and the MMF quark--propagator{} {\it Ansatz},
respectively.
In the case of the MMF quark propagator,
three different methods of calculation
(detailed in the text) yielded the same results.
The black dashed line represents the result
of Mello {\it et al.}~\cite{Mello:2017mor}.
The black solid line corresponds to the perturbative QCD result
(\ref{emff:FpiAsym}) with asymptotic PDA.}
\label{QuarkPropagators:3RQP:emff}
\end{figure}

  (a) $F_\pi(Q^2)$, given by Eq.~(\ref{Pi-EM-form-factor(4)}),
is calculated using FeynCalc and {\footnotesize PACKAGE-X} {\it Mathematica} packages
analogously to the $f_\pi$ calculation. The results are represented in
Fig.~\ref{QuarkPropagators:3RQP:emff}.

\begin{figure}[!h]
\centerline{\includegraphics[width=8cm,angle=0]{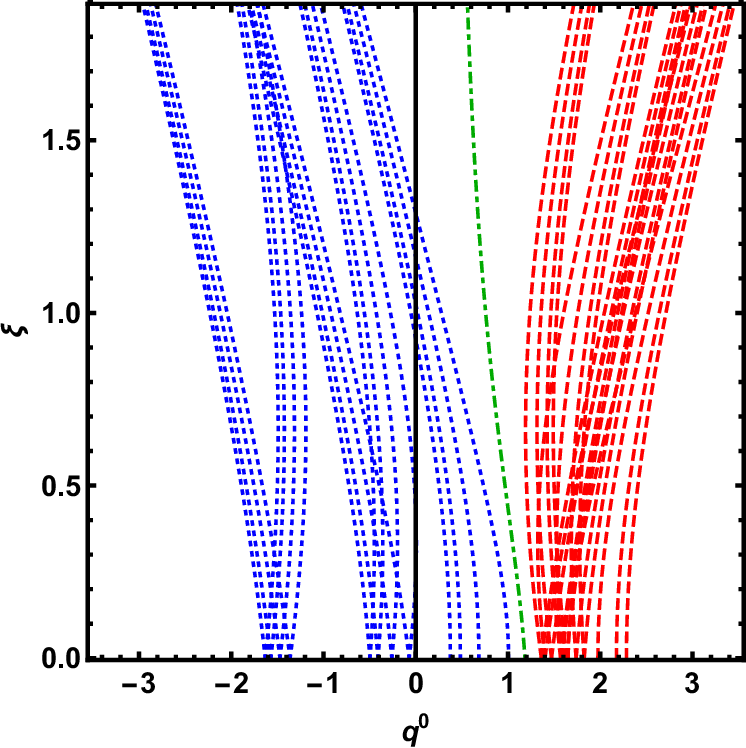}}
\caption{$(q^0)_j$'s and $(q^0)_c$ vs.
$\xi$ for $\vartheta=\pi/3$ and $Q^2=7~\unit{\GeV}^2$.
All in units of GeV.
Dot--dashed green line represents $(q^0)_c$,
blue dotted lines represent odd-indexed poles (set ${\cal A}$),
and red dashed lines represents even--indexed poles (set ${\cal B}$).
}
\label{QuarkPropagators:Mello-emff-path}
\end{figure}

  (b) Numerical integration is performed using adaptive quad\-ra\-ture:
expressing the space part of the four-vector $q$ in spherical coordinates,
$q=(q^0,\xi\sin\vartheta\cos\varphi,\allowbreak \xi\sin\vartheta\sin\varphi,\allowbreak \xi\cos\vartheta)$,
the poles of the integrand, in variable $q^0$, are
\begin{subequations}
\begin{align}
(q^0)_{1,2} & = \mp\sqrt{M_q^2+\xi^2-\xi\sqrt{Q^2}\cos\vartheta+Q^2/4}~,
\\
(q^0)_{3,4} & = \mp\sqrt{M_q^2+\xi^2+\xi\sqrt{Q^2}\cos\vartheta+Q^2/4}~,
\\
(q^0)_{5,6} & = 
   \frac{1}{2}\left(\sqrt{4M_\pi^2+Q^2}\mp 2\sqrt{M_q^2+\xi^2}\right)~,
\\
(q^0)_{7,8} & = 
   \frac{1}{4}\bigg(\sqrt{4M_\pi^2+Q^2}
\nonumber \\
   & \mp \sqrt{16 M_q^2+16\xi^2+8\xi\sqrt{Q^2}\cos\vartheta+Q^2}\bigg)~,
\\
(q^0)_{9,10} & = 
   \frac{1}{4}\bigg(\sqrt{4M_\pi^2+Q^2}
\nonumber \\
   & \mp \sqrt{16 M_q^2+16\xi^2-8\xi\sqrt{Q^2}\cos\vartheta+Q^2}\bigg)~,
\end{align}
\end{subequations}
where $M_q^2\in\{\mathfrak{p_1},\mathfrak{p_2},\mathfrak{p_3},\lambda^2\}$.
The numbers $(-M_q^2)$ are poles of the propagator functions (\ref{MMFQP})
and (\ref{QuarkPropagators:MelloMeloFredericoQP:MEu}). 
Changing $M_q^2\to M_q^2-i\varepsilon$ pushes odd--indexed poles
to the complex upper half--plane
and even-indexed poles to the lower half--plane.
We define two sets,
\begin{subequations}
\begin{align}
{\cal A} &= \Big\{ (q^0)_j\Big| j=1,3,5,7,9 \land M_q^2=\mathfrak{p}_1,\mathfrak{p}_2,\mathfrak{p}_3,\lambda^2 \Big\}~,
\\
{\cal B} &= \Big\{ (q^0)_j\Big| j=2,4,6,8,10 \land M_q^2=\mathfrak{p}_1,\mathfrak{p}_2,\mathfrak{p}_3,\lambda^2 \Big\}~,
\end{align}
\end{subequations}
where ${\cal A}$ and $\cal B$ contain poles that must be bypassed
from below and from above, respectively.
Note that not all four values of $M_q^2$ produce poles of the integrand.
For example, $(q^0)_{7,8}$ are poles of the integrand
only for $M_q^2=\lambda^2$; these two poles correspond to singular behavior
of $\Gamma_\pi(q-P/2,P^\prime)$ and are defined by equation
$(q-P/2)^2=\lambda^2$.
For simplicity of definition, sets ${\cal A}$ and ${\cal B}$ are allowed to
contain superfluous points, but this does not obstruct 
the analysis hereafter.
Numerical examination shows that $\max\{{\cal A}\}<\min\{{\cal B}\}$
for the chosen model parameters,
so we define
\begin{equation}
(q^0)_c = \frac{1}{2}\left(\max\{{\cal A}\} + \min\{{\cal B}\}\right)~,
\end{equation}
which is a function of $\vartheta$ and $\xi$, but does not depend
on $\varphi$ thanks to the symmetry.
Figure ~\ref{QuarkPropagators:Mello-emff-path} illustrates
the $\xi$ dependence of $(q^0)_j$'s and $(q^0)_c$ for a fixed value of
$\vartheta$.
Unlike the case of the $f_\pi$ calculation  (\ref{fpi_approx}),
where the first and third quadrants of the $q^0$ complex plane is free of
poles and the naive Wick rotation $q^0=-i q_4$ ($q_4\in\RealNumbers$)
is allowed, in the present case of the $F_\pi(Q^2)$ calculation,
the path of integration ought to be shifted to pass between poles contained in
the ${\cal A}$ and ${\cal B}$,
\begin{equation}
q^0 = (q^0)_c - i q_4~,
\label{ModifiedWickRotation}
\end{equation}
where $q_4\in\langle-\infty,\infty\rangle$.
Eventually, the numerical integration over $q_4$, $\xi$, and $\vartheta$
is performed using the adaptive quadrature;
see Fig.~\ref{QuarkPropagators:3RQP:emff}
for the final result.

  (c) Minkowski space integration utilizing light-cone momenta
is again performed analogously to the $f_\pi$ calculation. 
Now, there are 11
poles, in variable $q_-$, of the integrand of Eq.~(\ref{Pi-EM-form-factor(4)}).
The residua are calculated analytically and
adaptive quadrature is used for the final three--dimensional
integration.

   To conclude about the EMFF obtained with the MMF {\it An\-satz},
there are only insignificant differences, of order $\lesssim 0.1\%$,
between results for $F_\pi(Q^2)$
calculated using methods (a), (b), and (c).
The differences are compatible with the precision
of numerical integration that we prescribed in methods (b) and (c).
However, there is a significant discrepancy
between our results (blue dots) and those of Ref.~\cite{Mello:2017mor}
(black dashed line in our Fig.~\ref{QuarkPropagators:3RQP:emff}).
The MMF {\it Ansatz} \cite{Mello:2017mor} is also used 
in Ref.~\cite{deMelo:2019uho}, with the same model parameter values.
While $Q^2 F_\pi(Q^2)$ is
practically constant for $Q^2\gtrsim 3~\unit{\GeV}^2$ in the former paper, it
falls with $Q^2$ very noticeably in the latter one. 
Hence, Ref.~\cite{deMelo:2019uho} agrees better with our EMFF,
although it still falls more slowly than ours.

   For the ADFM quark propagator, we have calculated
$F_\pi(Q^2)$ using only one method out of three adopted 
for the MMF {\it Ansatz}; namely, method (b),
the modified Wick rotation, defined
by Eq.~(\ref{ModifiedWickRotation}), and subsequent three--di\-mension\-al 
adaptive Monte Carlo integration.
The results are depicted as red solid circles
in Fig.~\ref{QuarkPropagators:3RQP:emff}.

   Concerning the low--$Q^2$ behavior, the pion charge radius
$r_\pi=\sqrt{-6 F_\pi^\prime(0)}$
is calculated to
be $r_\pi=0.632~\fm$ and $0.699~\fm$ for MMF{} and ADFM{}
{\it Ans\"atze}, respectively.
Both values are reasonably near the experimental value of
$r_\pi=(0.659 \pm  0.004)~\fm$ \cite{ParticleDataGroup:2020ssz}.
The simple constituent quark model formula $r_\pi=\sqrt{3}/(2\pi f_\pi)$
\cite{Tarrach:1979ta,Gerasimov:1978cp} gives
$r_\pi=0.621~\fm$ and $0.760~\fm$
for MMF{} and ADFM{} {\it Ans\"atze}, respectively.
The approximate BS vertex (\ref{GTrelation})
does not guarantee
that the normalization condition $F_\pi(0)=1$ will be fulfilled.
The general form of the pseudoscalar BS vertex is
\begin{align}
& \Gamma_\pi(q,P)
=\gamma_5  \Bigl( H_1(q,P)
\nonumber \\ & + \slashed{P} H_2(q,P)
 + \slashed{q} H_3(q,P) + [\slashed{P},\slashed{q}] H_4(q,P) \Bigr)~,
\label{Gamma-parametrization}
\end{align}
where $H_1$, $H_2$, $H_3$, and $H_4$ are Lorentz--scalar 
functions \cite{LlewellynSmith:1969az}.
Solely keeping $H_1$ component and neglecting others,
just as we do in Eq.~(\ref{GTrelation}),
leads to deviation from $F_\pi(0)=1$ normalization condition
\cite{Maris:1997hd}.
We obtain $F_\pi(0)=0.950$ and $1.32$
for MMF{} and ADFM{} {\it Ans\"atze}, respectively
[which is interesting to compare, but of course
we could also follow Ref.~\cite{Mello:2017mor}, which
forces $F_\pi(0)=1$ by adjusting the normalization
of BS vertex (\ref{Gamma-parametrization});
for more details see Appendix \ref{sec:BSA_normalization}.]

The high--$Q^2$ asymptotics of the charged pion EMFF is discussed in
Sec.~\ref{sec:PDA} along with the asymptotics of the neutral pion TFF,
which is introduced in the next section.

\section{Transition form factor}
\label{sec:TFF}

\begin{figure}[t]
\centerline{\includegraphics[width=9cm,angle=0]{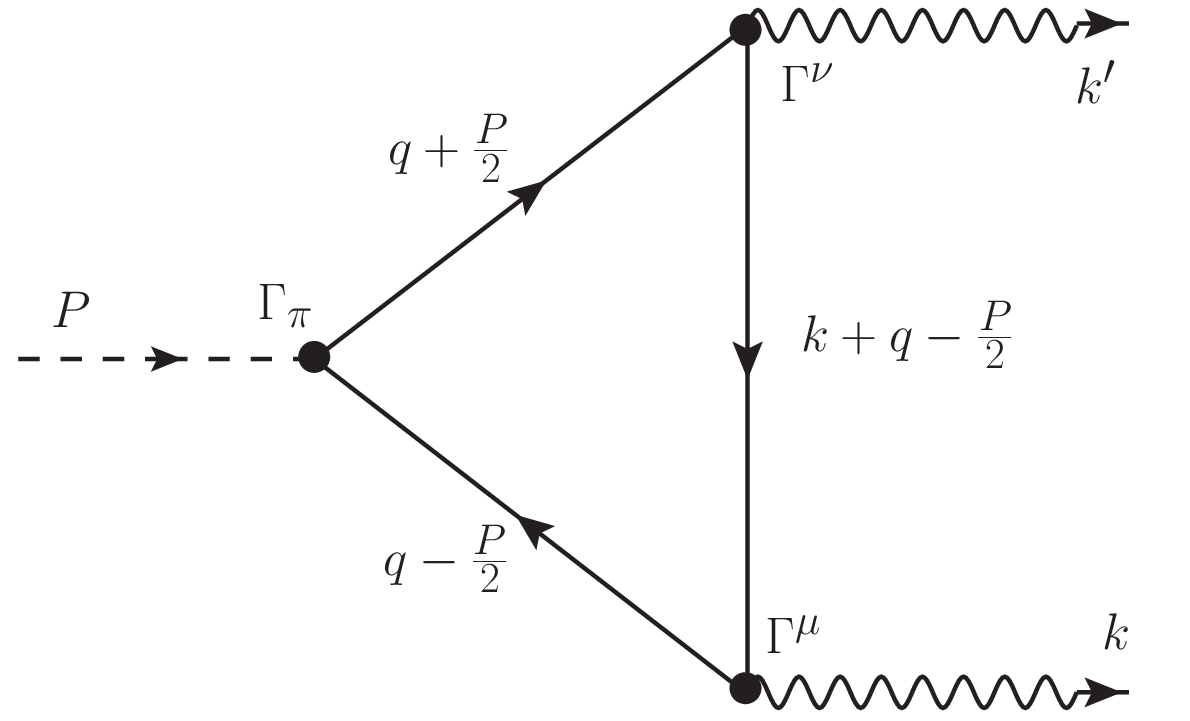}}
\caption{The quark triangle diagram for the transition
form factor calculation.}
\label{fig:Pi02gamma-3R}
\end{figure}

The two--photon amplitude $T(k^2,k^{\prime 2})$ 
that describes $\pi^0\to\gamma\gamma^{(\star)}$ processes,
depicted in Fig.~\ref{fig:Pi02gamma-3R}, is given by
        \begin{align}
        & T^{\mu\nu}(k,k^\prime)
	= \varepsilon^{\mu\nu\lambda\sigma} k_\lambda k^\prime_\sigma
        T(k^2,k^{\prime 2})
        \nonumber \\
        &=
        -
        N_c \,
        \frac{\mathcal{Q}_u^2-\mathcal{Q}_d^2}{2}
        \int\frac{d^4q}{(2\pi)^4} \mbox{\rm tr}  \Bigl\{
        \Gamma^\mu(q-\frac{P}{2},k+q-\frac{P}{2})
        \nonumber \\
        & \times
        S(k+q-\frac{P}{2})
        \Gamma^\nu(k+q-\frac{P}{2},q+\frac{P}{2})
        S(q+\frac{P}{2})
	\nonumber \\ & \times
        \Gamma_\pi(q,P)
        S(q-\frac{P}{2}) \Bigr\}
        + (k\leftrightarrow k^\prime,\mu\leftrightarrow\nu)~,
        \label{Tmunu(3)}
        \end{align}
in the GIA \cite{Roberts:1994hh,Frank:1994gc,Kekez:1998rw},
where $k$ and $k^\prime$ are the external photon momenta,
$P=k+k^\prime$ is the neutral pion momentum, and $P^2=M_\pi^2$.
The TFF is defined as
\begin{equation} F_{\pi\gamma}(Q^2)=|T(-Q^2,0)|~,\end{equation}
such that the $\pi^0\to\gamma\gamma$ decay width can be written as
\begin{equation}
   \Gamma(\pi^0\to\gamma\gamma)
   =
   \frac{\pi\alpha^2 M_\pi^3}{4}
  F_{\pi\gamma}(0)^2~.
  \label{pi2gg:GammaPi02gamma}
\end{equation}
 
\begin{figure}
\begin{center}
\includegraphics[width=9cm,angle=0]{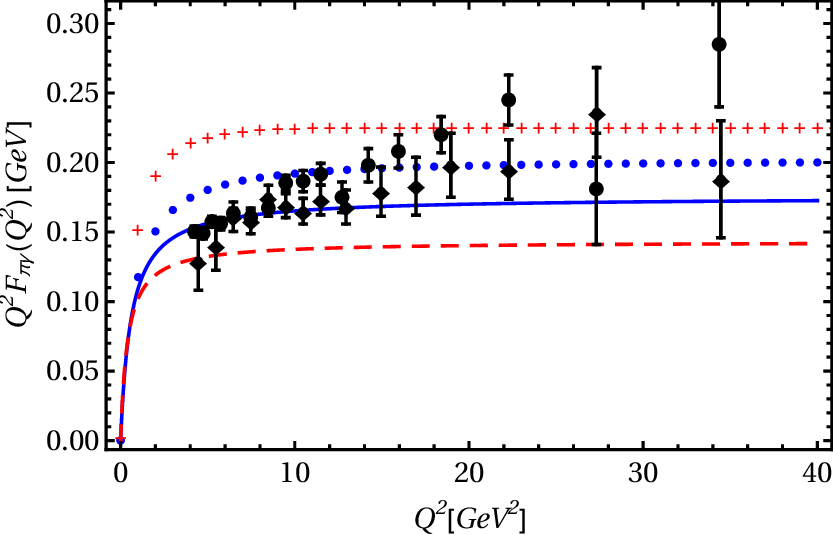}
\end{center}
\caption{Blue dots represent $\pi^0$ transition form factor calculated using
the MMF quark--propagator {\it Ansatz},
Eqs.~(\ref{QuarkPropagatorParametrization})
and (\ref{QuarkPropagators:MelloMeloFredericoQP:MandA}).
The red pluses are calculated using 
the ADFM quark propagator, Eqs.~(\ref{3RQP}).
The blue solid line and red dashed line represent the Brodsky--Lepage
interpolation formula (\ref{BrodskyLepage}), for the MMF quark--propagator{} and
ADFM quark--propagator{} models, respectively.
Solid circles and diamonds (with error bars) represent
the measurements of BaBar \cite{Aubert:2009mc} and Belle \cite{Uehara:2012ag}
Collaborations, respectively.
}
\label{fig:TFF-Mello+3R}
\end{figure}

   In respect of the MMF {\it Ansatz}, the
FeynCalc package is used to express the loop integral
in Eq.~(\ref{Tmunu(3)}) as a sum of the Passarino--Veltman functions,
while {\footnotesize PACKAGE-X} is used for the final numerical evaluation,
in a close analogy to the $F_\pi(Q^2)$ calculation, Sec.~\ref{sec:EFF},
method (a). The results of our calculation are pictorially
represented by the blue dots in Fig.~\ref{fig:TFF-Mello+3R}.
The experimental results are shown as solid
circles and diamonds (with error bars) in the same figure.

   On the other hand, the case of the ADFM {\it Ansatz} is treated
solely using method (b) described in Sec.~\ref{sec:EFF}.
The integrand appearing in Eq.~(\ref{Tmunu(3)}), as a function of $q^0$,
exhibits the same structure of the pole trajectories in the $q^0\xi$--plane,
as those illustrated in Fig.~\ref{QuarkPropagators:Mello-emff-path}
in the case of EMFF calculation.
The results are represented by the red pluses in Fig.~\ref{fig:TFF-Mello+3R}.

  It has been shown in Refs.~\cite{Bando:1993qy,Roberts:1994hh} that
the GIA amplitude (\ref{Tmunu(3)}) gives
$F_{\pi\gamma}(0)=1/(4\pi^2 f_\pi)$ in the chiral limit,
regardless of the specific choice of the quark dressing functions
$\sigma_V$ and $\sigma_S$, and in agreement with the
Adler--Bell--Jackiw (ABJ) anomaly result \cite{Adler:1969gk,Bell:1969ts}.
Our numerical results for $F_{\pi\gamma}(0)$ complies fairly
to this limit; the deviations are about 4.3\% and 0.7\%
for MMF and ADFM {\it Ans\"{a}tze}, respectively.

   The function $F_{\pi\gamma}(Q^2)$ is expected to be a smooth function 
near $Q^2=0$, down to $Q^2=-M_V^2$ where the vector--meson resonance peaks
appear; $V=\rho,\omega,\phi,\dots$.
The slope parameter $a$ is defined through the expansion
of the (normalized) TFF,
\begin{equation}
\frac{F_{\pi\gamma}(Q^2)}{F_{\pi\gamma}(0)}
=1-a\,\frac{Q^2}{M_\pi^2}+{\cal}O((Q^2)^2)~.
\end{equation}
The recent experimental result of the
NA62 Collaboration is $a=0.0368\pm 0.0057$ \cite{TheNA62:2016fhr};
the A2 Collaboration at MAMI gives $a=0.030\pm 0.010$ \cite{Adlarson:2016ykr}.
In both experiments, the Dalitz decay $\pi^0 \to e^+e^-\gamma$ is measured
for low timelike momentum transfer:
$-M_\pi^2\le Q^2=(p_{e^-}+p_{e^+})^2 \le -4m_e^2$.
Our calculation gives
$a=-M_\pi^2\,F_{\pi\gamma}^\prime(0)/F_{\pi\gamma}(0)=0.027$
for MMF {\it Ansatz} and $a=0.025$ for ADFM {\it Ansatz},
in reasonable agreement with the experimental values.
The following method was used to determine $a$.
We calculated several $(Q^2,F_{\pi\gamma}(Q^2))$ points
in the interval $-0.3~\unit{\GeV}^2\le Q^2\le 0.3~\unit{\GeV}^2$
and $-0.2~\unit{\GeV}^2\le Q^2\le 0.2~\unit{\GeV}^2$
for MMF and ADFM {\it Ansatz}, respectively.
These points were fitted to $F_{\pi\gamma}(Q^2)=A/(1+Q^2/B^2)$ curve;
the derivative $F_{\pi\gamma}^\prime(0)$ was computed from this fit.
A simple quark triangle model \cite{Ametller:1983ec} gives 
$a={M_\pi^2}/(12 M_c^2)$, where $M_c$ is the constituent quark mass.
Using $M_\pi=135~\unit{\MeV}$ and $M_c=M(0)=280~\unit{\MeV}$
(estimated from Fig.~\ref{fig:Mello:lattice}) gives $a=0.02$,
somewhat below the experimental values and our model results.
The high--$Q^2$ asymptotics of $F_{\pi\gamma}$ is addressed in
the next section and is compared with those calculated from the PDA.

\section{Pion distribution amplitude and asymptotics of form factors}
\label{sec:PDA}

\begin{figure}[h]
\centering
\includegraphics[width=9cm,angle=0]{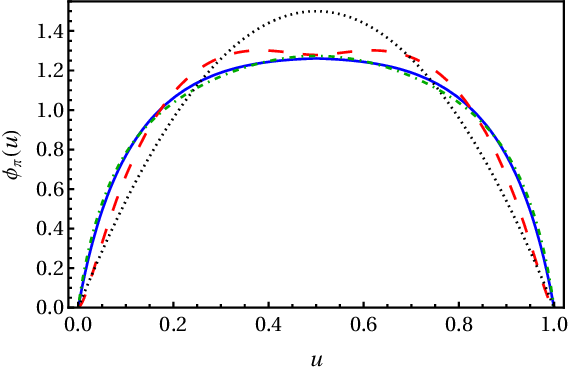}
\caption{Pion distribution amplitudes $\phi_\pi(u)$. 
Blue solid line and red dashed line correspond to
the MMF and ADFM {\it Ans\"{a}tze}, respectively.
Black dotted line represents the asymptotic form,
$\phi_{\pi}^{\mbox{\rm\scriptsize as}}(u) = 6 u(1-u)$.
Dash-dotted green line (very close to the solid blue one,
and hardly discernible from it) is the PDA from the state-of-the-art
SDE pion bound state, Eq. (22) in Ref. \cite{Roberts:2020hiw}.
\label{fig:pda}
}
\end{figure}

  The factorization property of the QCD hard scattering amplitudes
enables us to express these amplitudes
in terms of the pertinent distribution amplitudes.
The PDA, relevant for the TFF and EMFF calculation at large $Q^2$,
can be expressed as the light--cone projection,
\begin{equation}
\phi_\pi(u)
=
i\frac{N_c}{8\pi f_\pi}
\tr\left(\gamma_+ \gamma_5 
\int\frac{dq_{-}}{2\pi} 
\int\frac{d^2q_\perp}{(2\pi)^2} \, \chi_\pi(q,P) \right)~,
\label{LightFrontQuantization:PDAfromBSA}
\end{equation}
%
of the BS amplitude
\begin{equation}
\chi_\pi(q,P) = S(q+\frac{P}{2}) \Gamma_\pi(q,P) S(q-\frac{P}{2})
\end{equation}
\cite{Brodsky:1989pv,Brodsky:1997de,Lepage:1980fj,Chang:2013epa,Chang:2013pq}.
The variable $q_+$, which is implicit in the integrand of
Eq.~(\ref{LightFrontQuantization:PDAfromBSA}),
is defined by $u=1/2+q_+/P_+$.
The integral resembles
those of the $f_\pi$--calculation
(\ref{fpi_approx}) and could be treated in
the same way. For both propagator {\it Ans\"atze} we use
the Euclidean space integration, referred as method (b)
in Secs.~\ref{sec:PionDecayConstant} and \ref{sec:EFF}.
The resulting PDAs are displayed in Fig.~\ref{fig:pda}.

  The leading twist pQCD results for the asymptotics
of the pion form factor is
\cite{Farrar:1979aw,Efremov:1978rn,Efremov:1979qk,Lepage:1979zb}
        \begin{equation}
        F_\pi(Q^2)
        \sim
        \frac{16\pi\alpha_s(Q^2)f_\pi^2}{Q^2} 
	\left|\frac{1}{3} \int_0^1 du\, \frac{\phi_\pi(u)}{u} \right|^2
        \label{emff:FpiAsym}
        \end{equation}
for $Q^2\to\infty$,
where $\alpha_s$ is the QCD  running coupling constant:
$\alpha_s(Q^2) = {d\pi}/{\ln(Q^2/\Lqcd^{2})}$ at the one--loop order of
perturbation theory,
while $d$ is the same as in Eq.~(\ref{MEu:asymptotics}).
The renormalization scale ($\mu$) dependence of PDA is implicit here.
The asymptotic form of PDA,
$\phi_{\pi}^{\mbox{\rm\scriptsize as}}(u)=
\lim_{\mu\to\infty}\phi_\pi(u)\allowbreak =\allowbreak 6 u(1-u)$,
gives $\frac{1}{3}\int du\,\phi_\pi^{\mbox{\rm\scriptsize as}}(u)/u=1$,
leading to $ F_\pi(Q^2) \sim {16\pi\alpha_s(Q^2)f_\pi^2}/{Q^2}$
asymptotic behavior.
The PDAs $\phi_\pi(u)$, related to the models under consideration, do not
deviate too much from the asymptotic $\phi_\pi^{\mbox{\rm\scriptsize as}}(u)$
function; see Fig.~\ref{fig:pda}. The actual values of integrals are
$\frac{1}{3}\int du\,\phi_\pi(u)\allowbreak/u =$ $1.15$ and $1.02$  
for the MMF and ADFM models, respectively.
This results in respective 32\% and 4\% enhancement of $F_\pi(Q^2)$
relative to value obtained with $\phi_\pi^{\mbox{\rm\scriptsize as}}$.

   The asymptotic form of EMFF (\ref{emff:FpiAsym}),
being dependent on $\alpha_s(Q^2)$, critically reflects the perturbative
nature of high--energy QCD. Our simple meromorphic {\it Ans\"atze}
(\ref{QuarkPropagators:MelloMeloFredericoQP:MandA}) and (\ref{3RQP}),
which do not comply with the exact QCD asymptotics (\ref{MEu:asymptotics}),
is not expected to reproduce the UV logarithmic behavior
of Eq.~(\ref{emff:FpiAsym}).
We computed $F_\pi(Q^2)$ up to $Q^2=40~\unit{\GeV}^2$ and indeed found no evidence
that the asymptotic behavior $F_\pi(Q^2)\propto 1/(Q^2\ln(Q^2))$ was reached,
for either of our models.
The presently available experimental data on $F_\pi(Q^2)$ are anyway well above
the pQCD predictions (\ref{emff:FpiAsym}),
as discussed in Ref.~\cite{Cloet:2013jya} in more detail.

   The same PDA (\ref{LightFrontQuantization:PDAfromBSA}) also determines
the leading term of the light--cone expansion of form factor
$F_{\pi\gamma}(Q^2)$ \cite{Lepage:1980fj,Brodsky:2011yv},
\begin{equation}
F_{\pi\gamma}(Q^2) \sim \frac{2 f_\pi}{3Q^2} \int_0^1 
\frac{du\,\phi_\pi(u)}{(1-u)}~.
\label{QP:TFF:LeadingTwist}
\end{equation}
The asymptotic form of the PDA
leads to $F_{\pi\gamma}(Q^2) \sim 2 f_\pi/Q^2$ for $Q^2\to\infty$
asymptotic behavior \cite{Lepage:1980fj,Brodsky:1981rp}.
The Brodsky--Lepage (BL) dipole formula \cite{Brodsky:1981rp},
\begin{equation}
F_{\pi\gamma}(Q^2)
=
\frac{1}{4\pi^2 f_\pi}
\left( 1 + \frac{Q^2}{8\pi^2 f_\pi^2} \right)^{-1}~,
\label{BrodskyLepage}
\end{equation}
interpolates between $F_{\pi\gamma}(0)=1/(4\pi^2 f_\pi)$,
the ABJ anomaly result \cite{Adler:1969gk,Bell:1969ts},
and $\lim_{Q^2\to\infty} Q^2 F_{\pi\gamma}(Q^2)=2f_\pi$,
the pQCD limit.
The current experimental data \cite{Aubert:2009mc,Uehara:2012ag},
reaching up to $Q^2 \sim 35~\unit{\GeV}^2$,
do not show agreement with this limit yet.
On the theoretical side, recent SDE studies in Euclidean space
are not unanimous:
Raya {\it et al.} \cite{Raya:2015gva} are consistent with the hard scattering
limit, but Eichmann {\it et al.} \cite{Eichmann:2017wil} claim that
the BL limit is modified whenever the other external
photon is near on shell, {\it i.e}, $k^{\prime 2} \simeq 0$.
That is, some nonperturbative effects would always persist
in this case. The modified BL limit is also claimed independently from
the SDE approach, by some quite different theoretical studies
\cite{Hoferichter:2018kwz,Hoferichter:2020lap,Aoyama:2020ynm}.

As we can see from Fig.~\ref{fig:TFF-Mello+3R} and Table~\ref{tab:TFFLimits},
the high--$Q^2$ behavior of $F_{\pi\gamma}(Q^2)$
calculated in the GIA (\ref{Tmunu(3)}) deviates appreciably
from the BL limit of $2f_\pi/Q^2$ for both model {\it Ans\"atze}.
GIA limit of $F_{\pi\gamma}(Q^2)$ overshoots BL limit by 58\% and 15\% for
ADFM and MMF models, respectively.

\sisetup{round-mode = places, round-precision = 3}
\begin{table}[h]
\begin{ruledtabular}
\begin{tabular}{ccccc}
   approximation & ADFM & & MMF & \\
\hline
 \GIA{}         & \num{0.22596} & (\num{0.226306}) & \num{0.201817} & (\num{0.200065})
\\
 \bare{}        & \num{0.14545} & (\num{0.140809}) & \num{0.200458} & (\num{0.187294})
\\
 \BLnonasymptotic{} & \num{0.146308} & & \num{0.201797} &
\\
 \BL{}          & \num{0.14312} &  & \num{0.175120} &
\\
\end{tabular}
\end{ruledtabular}
\caption{
$\lim_{Q^2\to\infty} Q^2 F_{\pi\gamma}(Q^2)$, in units of GeV,
calculated using various approximation schemes.
The first two rows (denoted by GIA and {\it bare}), when related
to the ADFM {\it Ansatz}, are computed by fitting the function
$Q^2\mapsto \kappa_0 + \kappa_1/Q^2$ to a set of
discrete values of $Q^2 F_{\pi\gamma}(Q^2)$ calculated
in the interval $10~\unit{\GeV}^2\le Q^2\le 50~\unit{\GeV}^2$.
The symbols $\kappa_n$ $(n=0,1,2,3)$ denote the fitting constants.
In the MMF case, the corresponding limits are computed by fitting the function
$Q^2\mapsto \kappa_0 + \kappa_1/Q^2 + \kappa_2/(Q^2)^2 + \kappa_3/(Q^2)^3$
to values of $Q^2 F_{\pi\gamma}(Q^2)$ calculated
in the interval $20~\unit{\GeV}^2\le Q^2\le 100~\unit{\GeV}^2$.
Thus, $\lim_{Q^2\to\infty} Q^2 F_{\pi\gamma}(Q^2)=\kappa_0$
for both GIA and {\it bare} rows, for the both {\it Ans\"atze}.
In the brackets are the results of the same GIA and
{\it bare} calculations obtained at $\, Q^2 = 40\unit{\GeV}^2$, to
illustrate the differences between $\, Q^2 F_{\pi\gamma}(Q^2)$
at a large but finite $\, Q^2$ and in the $\, Q^2\to\infty$ limit.
The last two rows are calculated from Eq.~(\ref{QP:TFF:LeadingTwist}),
using the appropriate PDAs. Thus, the fourth row
	is simply $\, 2 f_\pi$ due to
$\phi_{\pi}(u) =\phi_{\pi}^{\mbox{\rm\scriptsize as}}(u)$.
	However, the third row differs from 
$\, 2 f_\pi$, since PDAs used in Eq.~(\ref{QP:TFF:LeadingTwist})
are not asymptotic, but calculated in the MMF and ADFM models
through Eq. (\ref{LightFrontQuantization:PDAfromBSA}).
}
\label{tab:TFFLimits}
\end{table}

   The row denoted by ``\bare{}'' in Table~\ref{tab:TFFLimits} is
calculated from Eq.~(\ref{Tmunu(3)}) by replacing the
dressed electromagnetic vertices $\Gamma^\mu(q,q^\prime)$ with the
bare ones $\gamma^\mu$ and the quark propagators
$S(l)$ that propagate hard momenta $l=q\pm(k-k^\prime)/2$,
with the bare (and massless) ones, $\slashed{l}/l^2$. 
This leads to a much simpler expression for $T^{\mu\nu}$:
\footnote{Compare this to our previous and a little bit cruder approximation
\cite{Kekez:1998rw,Klabucar:1998hr}. See also related
Refs.~\cite{Tandy:1998ha,Roberts:1998gs}. That approximation gave a
universal $T(-Q^2,-Q^{\prime 2})\sim (4/3)(f_\pi/(Q^2+Q^{\prime 2}))$ 
behavior for large $Q^2+Q^{\prime 2}$,
which was criticized in Ref.~\cite{Anikin:1999cx}.
}
\begin{align}
&T^{\mu\nu}(k,k^\prime)
=
-2iN_c\,\frac{\mathcal{Q}_u^2-\mathcal{Q}_d^2}{2}\,
\varepsilon^{\mu\nu\lambda\sigma}\,
\nonumber \\ &\times
\int\frac{d^4q}{(2\pi)^4}
\frac{\left[ \frac{1}{2}(k^\prime-k)-q \right]_\lambda }
     {\left[ \frac{1}{2}(k^\prime-k)-q \right]^2}\,
\tr\{ \gamma_\sigma \gamma_5 \chi_\pi(q,P) \}~.
\label{PionDistributionAmplitude:T_munu}
\end{align}
The pertaining $\lim_{Q^2\to\infty} Q^2 F_{\pi\gamma}(Q^2)$
deviates negligibly from the corresponding \GIA{} value
in the case of the MMF {\it Ansatz}, but the deviation is significant
in the case of the ADFM {\it Ansatz}.

  It is apparent that
not all quark legs attached to electromagnetic vertices
carry the large momentum scale $Q^2$
\footnote{Compare to Ref.~\cite{Roberts:2010rn},
Sec.~III.B.1, last paragraph.}
; see Fig.~\ref{fig:Pi02gamma-3R}.
It is enough to improve the previous \bare{} approximation
(\ref{PionDistributionAmplitude:T_munu}) such that we partially restore
these soft contributions originally present in the Ball--Chiu vertex
(\ref{Ball-Chiu-vertex}),
\begin{align}
\gamma^{\mu(\nu)}
\to\frac{1}{2}\left(1+\AEu\Bigl(-[q-(+)\frac{P}{2}]^2\Bigr)\right) \gamma^{\mu(\nu)}~,
\label{softEMvertex}
\end{align}
and the \GIA{} limit is recovered
\cite{Kekez:1998rw,Klabucar:1998hr},
\begin{equation}
\lim_{Q^2\to\infty} Q^2 
  F^{\mbox{\rm\scriptsize ``$(1+A_{\mbox{\rm\scriptsize soft}})/2$''}}_{\pi\gamma}(Q^2)
= 0.225~\unit{\GeV}~,
\end{equation}
where the superscript ``$(1+A_{\mbox{\rm\scriptsize soft}})/2$'' indicates
that $F_{\pi\gamma}$ is calculated using vertex (\ref{softEMvertex})
instead of the Ball--Chiu one.
Hence, the nontrivial infrared behavior of the wave function renormalization
$\ZEu(x)=1/\AEu(x)\ne 1$ is responsible for the two calculations,
the first one based on \GIA{} Eq.~(\ref{Tmunu(3)})
and the second one based on
\bare{} Eq.~(\ref{PionDistributionAmplitude:T_munu}),
producing unequal asymptotics of $F_{\pi\gamma}(Q^2)$.
Of course, for the MMF{} {\it Ansatz},
where $\ZEu(x)\equiv 1$, both calculations give the same asymptotic limit.

  The respective integral $\frac{1}{3}\int du\, \phi_\pi(u)/u$
values of 1.02 and 1.15 for the MMF and ADFM {\it Ans\"atze},
which influence the EMFF asymptotics (\ref{emff:FpiAsym}),
are reflected also in the asymptotic behavior of the TFF calculated
from Eq.~(\ref{QP:TFF:LeadingTwist})
and shown in Table~\ref{tab:TFFLimits}, in the row denoted by 
\BLnonasymptotic{} [for it is not calculated using the 
asymptotic form of $\phi(u)$, but the model calculated one].

   To the end of this section, we explain the similarity between
the \bare{} and \BLnonasymptotic{} approximation.
Light--cone expansion of the time--ordered product
of two electromagnetic currents,
$T\{J^\mu(x),J^\nu(y)\}$,
leads to the following approximate expression:
\begin{align}
&T^{\mu\nu}(k,k^\prime)
\simeq
2\frac{\mathcal{Q}_u^2-\mathcal{Q}_d^2}{\sqrt{2}}
\frac{1}{2\pi^2}
\varepsilon^{\mu\nu\lambda\sigma}
\nonumber \\ &\times
i\int d^4z\,e^{ik^\prime\cdot z}
\frac{z_\lambda}{z^4}
\langle\mbox{\rm vac}|
   :\bar{d}(0)\gamma_\sigma \gamma_5 u(z): |\pi^+(P)\rangle_{z^2=0}~.
\nonumber \\
\label{PionDistributionAmplitude:Tmunu:LightCone}
\end{align}
%
(See, {\it e.g.}, Refs.~\cite{Radyushkin:2000ap,Braun:2011dg}.)
\footnote{In the isospin limit
 $\sqrt{2}\langle\mbox{\rm vac}|
   :\bar{d}(0)\gamma_\sigma \gamma_5 u(z): |\pi^+(P)\rangle
\\
=
\langle\mbox{\rm vac}|
   :\bar{u}(0)\gamma_\sigma \gamma_5 u(z)
   -
    \bar{d}(0)\gamma_\sigma \gamma_5 d(z): |\pi^0(P)\rangle$.}
The path--ordered ``string operator,''
\begin{equation}
P \exp\left( ig\int_x^0 A^\alpha(y)dy_\alpha \right)~,
\label{WillsonLine}
\end{equation}
must be included between the quark fields. This operator equals
unity in light--cone gauge; see, {\it e.g.}, Ref.~\cite{Brodsky:1997de}.

  On the one hand, expressing the above $\pi^+$-to-vacuum matrix element
through the the BS amplitude,
\begin{align}
&\lvac : \bar{u}(0) \gamma^\mu \gamma_5 u(z)
         - 
         \bar{d}(0) \gamma^\mu \gamma_5 d(z) :
 |\pi^0(P)\rangle
\nonumber \\
&=
- N_c\,
e^{-iP\cdot z/ 2}
\int\frac{d^4q}{(2\pi)^4}
e^{-iq\cdot z}
\tr\left(\gamma^\mu\gamma_5\chi_\pi(q,P)\right)~,
\label{LightConeQuantization:VacA5muPi}
\end{align}
%
we reproduce \bare{} Eq.~(\ref{PionDistributionAmplitude:T_munu}).
On the other hand, the definition of the PDA,
\begin{align}
&\frac{1}{2}\lvac :\bar{u}(0) \gamma^\mu \gamma_5 u(z)
                -  
	 	   \bar{d}(0) \gamma^\mu \gamma_5 d(z) :
        |\pi^0(P)\rangle_{z_+=z_\perp=0}
\nonumber \\
&=
i \delta^{ab} f_\pi P^\mu \int_0^1 du\, e^{-iuP\cdot z}\, \phi_\pi(u)~,
\end{align}
%
leads eventually to the \BLnonasymptotic{} approximation,
Eq.~(\ref{QP:TFF:LeadingTwist}).
To conclude, both Eqs.~(\ref{QP:TFF:LeadingTwist}) and
(\ref{PionDistributionAmplitude:T_munu}) follow from
Eq.~(\ref{PionDistributionAmplitude:Tmunu:LightCone}),
except Eq.~(\ref{QP:TFF:LeadingTwist}) is derived without the
$z^2=0$ constraint, {\it i.e.}, without light-cone projection of the
nonlocal operator
${:\bar\psi(0) \frac{\lambda^a}{2} \gamma^\mu \gamma_5 \psi(z):}$.
It turns out that such a difference is of little influence, at least
for the models under consideration. 

\section{Summary and conclusions}
\label{sec:SummaryAndConclusions}

   In this paper, we have studied
two meromorphic {\it Ans\"atze} for the dressed quark propagator
(suggested in Refs.~\cite{Mello:2017mor} and \cite{Alkofer:2003jj}),
which represent strongly nonperturbative dressing, but still permit
formulating clear connections between Euclidean and Minkowski spacetime
calculations.
Thanks to the quark--level Goldberger--Treiman relation 
(\ref{GTrelation}),
the pseudoscalar BS vertex can be related to the dynamically
dressed mo\-men\-tum--dependent quark mass function \cite{Roberts:1994dr}.
Additionally, by exploiting the Ball--Chiu vertex
\cite{Ball:1980ay,Ball:1980ax}
as an approximation for the fully dressed quark--quark--photon vertex,
we are provided with all the necessary elements to calculate
the pion decay constant, EMFF, TFF, and PDA.
The related amplitudes were calculated using several methods
in order to check the robustness of the results.

   The used quark {\it Ans\"atze} as well as the pertaining vertices
exhibit masslike singularities
on the real timelike mo\-men\-tum axis and do not
obey the pQCD asymptotic behavior; hence, we can hardly expect that
the correct perturbative asymptotic behavior of the electromagnetic form factor
$F_\pi(Q^2)\propto 1/(Q^2\ln(Q^2))$ will be attained. Indeed,
our numerical evaluation of $F_\pi(Q^2)$ up to $Q^2=40~\unit{\GeV}^2$ did not
show evidence that either $F_\pi(Q^2)\sim 1/(Q^2\ln(Q^2))$ limit or 
simpler power--law $F_\pi(Q^2)\allowbreak\sim 1/Q^2$ limit is reached.
However, it should be acknowledged that
the exact asymptotic behavior is of purely academic interest here
because (a) it is generally expected that the asymptotic regime probably
starts at $Q^2\gtrsim 20~\unit{\GeV}^2$,
well above Jefferson Lab capability after proposed upgrade \cite{PR12-06-101},
(b) and even existing Cornell experimental data at $Q^2=6.30~\unit{\GeV}^2$
and $9.77~\unit{\GeV}^2$ have large error bars \cite{Bebek:1977pe}.
For high $Q^2$, our results for $Q^2 F_\pi(Q^2)$ obviously deviate
from those of Ref.~\cite{Mello:2017mor}.
The low--$Q^2$ behavior of $F_\pi(Q^2)$,
encoded in the pion charge radius $r_\pi$,
was found to be in a reasonable agreement with experiment,
given the simplicity of the model.

   The leading--order pQCD expression for the high--$Q^2$ behavior
of the transition form factor, $F_{\pi\gamma}(Q^2)\sim 2f_\pi/Q^2$,
depends only on $f_\pi$, the low--energy pion observable, which is pretty
insensitive to the details of the high--energy dynamics.
Hence, we could naively expect that our {\it Ans\"atze},
despite not incorporating the exact perturbative regime behavior,
should produce the correct perturbative limit 
of the pion transition form factor.
However, in the generalized impulse approximation
the electromagnetic vertices keep one quark leg soft,
even for the high--$Q^2$ external photon.
As the result, this approximation gave
$Q^2 F_{\pi\gamma}(Q^2)$ finite for $Q^2\to\infty$,
but, similar to Refs. \cite{Eichmann:2017wil,Hoferichter:2018kwz,Hoferichter:2020lap,Aoyama:2020ynm},
generally unequal to the pQCD limit of $2 f_\pi$;
see also Refs.~\cite{Kekez:1998rw,Klabucar:1998hr}.
In relation to low--$Q^2$ behavior,
our results for the TFF slope parameter are 10\%-15\%
below the experimental value.

   The pion distribution amplitudes that were calculated using our
{\it Ans\"{a}tze} did not deviate appreciably from the asymptotic one.
If we input these amplitudes (instead of the asymptotic one)
to the pQCD form--factor formulae, the result is enhanced
up to ~30\%, depending on the form factor and {\it Ans\"atze}.
When one compares the results given above, the MMF {\it Ansatz}
is considerably more successful than the ADFM one.
In part, this can be explained by noticing that the PDA that
we obtained from the MMF {\it Ansatz} (the solid curve in Fig. \ref{fig:pda})
is very close to the PDA calculated from the pion bound-state amplitude
obtained using the most sophisticated SDE kernel \cite{Roberts:2020hiw}.
This PDA is given by Eq. (22) in Ref. \cite{Roberts:2020hiw}, and 
is hardly discernible from our solid curve in Fig. \ref{fig:pda}.
This indicates that the MMF {\it Ansatz} at least partially captures the results obtained
by some of the presently most advanced SDE calculations \cite{Roberts:2020hiw}.
The MMF {\it Ansatz} (\ref{QuarkPropagators:MelloMeloFredericoQP:MandA}) is more
realistic also in that it incorporates the explicit chiral symmetry breaking,
whereas the ADFM one (\ref{3RQP}) corresponds to the chiral limit,
for which the ADFM paper \cite{Alkofer:2003jj} concludes that their 
parametrizations should yield values of the pion decay constant
$f_\pi$ 10-20\% below the empirical value. Their value is thus just
$f_\pi = (71 \pm 3)$ MeV \cite{Alkofer:2003jj}, obtained for the presently
adopted ADFM {\it Ansatz} and its parameters in Eq. (\ref{3RQP}). Hence,
the large difference between MMF and ADFM results for $f_\pi$ is
inherited from the respective Refs. \cite{Mello:2017mor,Alkofer:2003jj},
since we adopted their respective
{\it Ans\"atze} and parameters without change.

In the present paper, however, more important than the
phenomenological considerations is the following:
the simple analytic structure of quark--propagator {\it Ans\"{a}tze}
employed, together with suitable approximations for the required vertices,
enabled us to keep control of the Wick rotation when calculating some processes;
the pertinent amplitudes can be calculated equally
in Minkowski and Euclidean space.
Kindred studies are mostly restricted to the Euclidean space;
their propagators and vertices are sensibly defined for
spacelike external momenta,
$q^2 = (q^0)^2-|\vect{q}|^2<0$,
but their analytic properties
(singularities in the first and third quadrants of the complex $q^0$ plane)
preclude Wick rotation back to the Minkowski space. In principle,
it is not difficult to impose the correct perturbative asymptotic
behavior on gluon and  quark propagators in such models.
In the context of the coupled Schwinger--Dyson and Bethe--Salpeter
equation, such an example is provided in
Refs.~\cite{Jain:1991pk,Munczek:1991jb,Jain:1993qh}; a similar and widely used
model is introduced in Refs.~\cite{Maris:1997tm,Maris:1999nt} and
its application reviewed in Ref.~\cite{Maris:2003vk}.
Among the variety of quark--propagator {\it Ans\"atze} explored in
Ref.~\cite{Alkofer:2003jj} that exhibit correct pQCD behavior,
none is suitable for the calculation methods presented in this work:
the branch cut in propagator functions do not allow the use
of perturbative techniques
while the complicated singularity structure prevents the Wick rotation.

  Future work may include calculation of some other processes involving
quark loops, {\it e.g.}, $\gamma^\star\to 3\pi$, $\gamma\gamma\to\pi\pi$,
and $\pi^0\to e^-e^+$.
The most appealing improvement would be a quark propagator
{\it Ansatz} that has the correct UV behavior
and, at the same time, enough simple analytic structure that allow 
Wick rotation (in the sense used in this paper).
But it is not evident to us whether such a task could be achieved.

\section*{Acknowledgments} 

  This work was supported in part by
by STSM grants from COST Actions CA15213 THOR and CA16214 PHAROS.
The Feynman diagrams were drawn with the help of {\footnotesize JAXODRAW} \cite{Binosi:2003yf},
based on {\footnotesize AXODRAW} \cite{Vermaseren:1994je}.

\appendix

\section{Normalization of the BS amplitude}
\label{sec:BSA_normalization}

The matrix element of the electromagnetic current is generally
\begin{align}
\langle \pi^{+}(P^\prime)|J^\mu(x)|\pi^{+}(P)\rangle =
e^{-i(P^\prime-P)\cdot x}
\nonumber \\ \times
\big(
   (P^\mu + P^{\prime\mu}) F_\pi (Q^2)
   +
   (P^\mu - P^{\prime\mu}) G_\pi (Q^2)
\big)~.
\end{align}
The electromagnetic current conservation $\partial_\mu J^\mu(x)=0$
implies $G_\pi(Q^2)=0$.
Our pion states normalization,
\begin{equation}
\langle\pi^+(P^\prime)|\pi^+(P)\rangle
=
(2\pi)^3\, 2 E(\vect{P})\, \delta^{(3)}(\vect{P}-\vect{P}^\prime)~,
\label{pi-norm}
\end{equation}
where $E(\vect{P})=\sqrt{M_\pi^2+|\vect{P}|^2}$,
together with
\begin{equation}
\hat{Q} |\pi^{+}(P)\rangle 
= \int d^3x J^0(x) |\pi^{+}(P)\rangle
= |\pi^{+}(P)\rangle
\end{equation}
automatically ensures that
\begin{equation}
F_\pi(0)=1~.
\end{equation}
In the chiral limit, the axial--vector Ward--Takahashi identity reads
\begin{equation}
(p^\prime-p)_\lambda \Gamma_5^{\,\,a\lambda}(p^\prime,p)
=
\left(
S^{-1}(p^\prime) \gamma_5 + \gamma_5 S^{-1}(p) \right)
\frac{\lambda^a}{2}~.
\label{AV-WT}
\end{equation}
The pion pole contribution to the axial--vector vertex is
\begin{equation}
\Gamma_{5}^{\,\,a\lambda}(p^\prime,p)
\simeq
\frac{\lambda^a}{2} \,
f_\pi P^\lambda
\frac{\Gamma_\pi(q,\vect{P})} {P^2}~,
\label{Lambda_5^amu-EdenAlt}
\end{equation}
where $p=q-\frac{P}{2}$ and $p^\prime=q+\frac{P}{2}$.
This leads eventually to our Eq.~(\ref{GTrelation}),
\begin{equation}
\Gamma_\pi(q,P) \simeq
\Gamma_\pi(q,0) =  -\frac{2 (\BEu(-q^2))_{\mbox{\rm\scriptsize c.l.}}}{f_\pi} \gamma_5~.
\label{GTrelation-again}
\end{equation}
Equations (\ref{AV-WT}) and (\ref{Lambda_5^amu-EdenAlt}) fix normalization
of $\Gamma_\pi$ as it is given by Eq.~(\ref{GTrelation-again}).
If we plug the approximate $\Gamma_\pi$, Eq.~(\ref{GTrelation-again}),
into Eq.~(\ref{Pi-EM-form-factor(4)}) we can expect that the resulting 
$F_\pi(0)\ne 1$. Indeed, for {MMF} and {ADFM} {\it Ans\"atze} we
get $F_\pi(0)=0.950$ and $F_\pi(0)=1.32$, respectively. 
Deviation from $F_\pi(0)=1$ measures quality of the approximation 
(\ref{GTrelation-again}).
Alternatively, following Ref.~\cite{Mello:2017mor},
we could modify Eq.~(\ref{GTrelation-again}) by introducing
an additional normalization factor $\mathfrak{N}$,
\begin{equation}
\Gamma_\pi(q,P) \simeq  - \mathfrak{N}\, \frac{2 (\BEu(-q^2))_{\mbox{\rm\scriptsize c.l.}}}{f_\pi} \, \gamma_5~.
\label{GTrelation-norm}
\end{equation}
If we denote by $f_\pi^0$, $F_\pi^0$, $r_\pi^0$, and $F_{\pi\gamma}^0$
the quantities calculated using Eq.~(\ref{GTrelation-again}) and 
by $f_\pi$, $F_\pi$, $r_\pi$, and $F_{\pi\gamma}$ those calculated using
Eq.~(\ref{GTrelation-norm}), 
the relation between these two sets will be
\begin{subequations}
\begin{align}
f_\pi &= \sqrt{\mathfrak{N}}\, f_\pi^0~,
\\
F_\pi(Q^2) &= \mathfrak{N}^2\, F_\pi^0(Q^2)~,
\label{Fpi-norm}
\\
r_\pi &= \mathfrak{N}\, r_\pi^0~,
\\
F_{\pi\gamma}(Q^2) &= \sqrt{\mathfrak{N}}\, F_{\pi\gamma}^0(Q^2)~.
\end{align}
\end{subequations}
Then we could impose the constraint $F_\pi(0)=1$, calculate $\mathfrak{N}$
from Eq.~(\ref{Fpi-norm}) and relate $f_\pi$, $F_\pi(Q^2)$,
and $F_{\pi\gamma}(Q^2)$ to 
$f_\pi^0$, $F_\pi^0(Q^2)$, and $F_{\pi\gamma}^0(Q^2)$.

  See Refs.~\cite{Maris:1997hd,Maris:1997tm} about the relationship,
in the chiral limit, between the
normalization of the pion BS vertex and $f_\pi$.

\bibliography{funnel}

\end{document}